\documentclass[runningheads]{llncs}
\usepackage{flushend}
\usepackage{graphicx}
\usepackage[style=base]{caption}
\usepackage{subcaption}
\usepackage{authblk}

\usepackage{tikz,xcolor,hyperref}

\definecolor{lime}{HTML}{A6CE39}
\DeclareRobustCommand{\orcidicon}{
	\begin{tikzpicture}
	\draw[lime, fill=lime] (0,0) 
	circle [radius=0.16] 
	node[white] {{\fontfamily{qag}\selectfont \tiny ID}};
	\draw[white, fill=white] (-0.0625,0.095) 
	circle [radius=0.007];
	\end{tikzpicture}
	\hspace{-2mm}
	}
\foreach \x in {A, ..., Z}{%
	\expandafter\xdef\csname orcid\x\endcsname{\noexpand\href{https://orcid.org/\csname orcidauthor\x\endcsname}{\noexpand\orcidicon}}
}

\begin{document}

\title{Development of a Machine Learning Model and Mobile
Application to Aid in Predicting Dosage of Vitamin K
Antagonists Among Indian Patients   \thanks{The current work is the result of a Memorandum of Understanding between National Institute of Technology Calicut  and Sree Chitra Tirunal Institute for Medical Sciences and Technology, Thiruvananthapuram}}
\titlerunning{Machine learning model for anti-coagulation guidance}

% \author{Amruthlal M (1) \and
% Devika S (1)  \and
% Ameer Suhail P A (1) \and
% Aravind K Menon (1) \and
% Vignesh Krishnan (1) \and
% Alan Thomas (1) \and
% Manu Thomas (1) \and
% Sanjay G (2) \and 
% Lakshmi Kanth L.R. (2) \and  
% Jeemon P (2) \and 
% Jimmy Jose (1) \and
% Harikrishnan S (2)}
% %
% \authorrunning{Amruthlal M et al.}

% \institute{Department of Computer Science and Engineering,\\
% National Institute of Technology Calicut, India
% \email{jimmy@nitc.ac.in}\\
%  \and
% Sree Chitra Tirunal Institute for Medical Sciences and Technology, Thiruvananthapuram, India\\
% \email{drharikrishnan@outlook.com}}

\author{Amruthlal M\inst{1} \and Devika S\inst{1} \and Ameer Suhail P A\inst{1} \and Aravind K Menon\inst{1} \and Vignesh Krishnan\inst{1} \and Alan Thomas\inst{1} \and Manu Thomas\inst{1} \and Sanjay G \inst{2} \and Lakshmi Kanth L R \inst{2} \and Jimmy Jose\inst{1}\orcidA{} %\orcidID{0000-0001-7074-090X} 
\and Harikrishnan S\inst{2}}

% \institute{Department of Computer Science and Engineering,\\ Indian Institute of Technology Kharagpur, India\\
% \email{\{jimmy,drc\}@cse.iitkgp.ernet.in}
% \and Infinera India Pvt Ltd\\
% \email{sourav10101976@gmail.com}}
\institute{Department of Computer Science and Engineering,\\ National Institute of Technology Calicut, India\\
\email{jimmy@nitc.ac.in} 
\and Sree Chitra Tirunal Institute for Medical Sciences and Technology, Thiruvananthapuram, India\\
\email{drharikrishnan@outlook.com}}

\authorrunning{Amruthlal M et al.}
%\institute{Alcatel-Lucent India Ltd\\
%Email:sourav10101976@gmail.com}

%
\maketitle              % typeset the header of the contribution
\begin{abstract}
Patients who undergo mechanical heart  valve replacements or have conditions like Atrial Fibrillation have to take Vitamin K Antagonists (VKA) drugs to prevent coagulation of blood. These drugs have narrow therapeutic range and need to be very closely monitored due to life threatening side effects. The dosage of VKA drug is determined and revised by a physician based on Prothrombin Time - International Normalised Ratio (PT-INR) value obtained through   a blood test. Our work aimed at predicting the maintenance dosage of warfarin, the present most widely recommended anticoagulant drug, using the de-identified medical data collected from 109 patients from Kerala. A Support Vector Machine (SVM) Regression model was built to predict the maintenance dosage of warfarin, for patients who have been undergoing treatment from a physician and have reached stable INR values between 2.0 and 4.0.

\end{abstract}
\section{Introduction}
Patients who undergo cardiac procedures such as mechanical heart valve replacement and those who have atrial fibrillation require oral anticoagulant (OAC) drugs, mostly Vitamin K antagonists (VKA) to prevent blood clotting.  The dosage of VKA drugs are monitored by physicians by observing PT-INR (Prothrombin Time – International Normalised Ratio) values  obtained through a  blood test. These VKA drugs have got a very narrow therapeutic range\cite{warfarin_narrow}, so they require very close monitoring. This will normally require the services of a physician.

 \par Developing countries like India have disparities in access to healthcare where the patients find it difficult to get the services of a physician. Patients have to travel long distances to meet a physician to show the results of PT-INR and alter the dosage of the drugs, if necessary. The cost of travelling and physical disabilities, which prevent patients from travelling, make recurrent blood tests a tedious process. This forces the patient to avoid doing the test, which can lead to bleeding or formation of blood clots inside blood vessels (thrombosis).
 
 \par Of late,  the PT-INR test is widely available in many small laboratories even in villages and also as point-of-care (POC) devices, which are readily available and can be used at home. If we can predict the warfarin dosage from PT-INR results using a handheld device like a mobile phone or a computer based application, it will be useful to people with limited healthcare access.
 
\par Identifying a stable dose of warfarin just after initiation of the drug is a tedious process which requires supervision of a physician and is usually done before the patient is discharged, for example after the valve surgery \cite{warfarin_narrow}. Our work focuses on predicting the maintenance dosage in patients who are on follow-up at home with stable INR readings for some time.  

\par Studies  in the field have shown that, incorporating pharmacogenomic data can lead to  higher accuracy compared to clinical models in warfarin dosage prediction \cite{ref_article4}. But the genetic information used in the pharmacogenetic models is difficult to obtain and the gene variant with biggest influence on prediction varies between populations \cite{ref_article2}. So we thought of developing a simple algorithm based on previous and current INR values and the last drug dose, as required for the different indications. 

 \par This paper is organized as follows. The next section, Section 2 discusses about existing literature on clinical and pharmacogenetic warfarin dosage prediction. Details of preliminary analysis of dataset used for the work is provided in Section 3. Section 4 discusses different machine learning models that were tried out. Section 5 describes the algorithm used to convert predicted daily dosage (in decimals) to weekly dosage sequence (in integer) to match current clinical practice. Section 6 is about the client application designed to be used by patients and doctors for dosage prediction.

% \cite{ref_article1,ref_article2,ref_article3,ref_article4,warfarin_demography,warfarin_initial_dosage,warfarin_novac,warfarin_switch,one_hot,svm}.

\section{Literature Survey on Methods of
Warfarin Prediction}

\par As mentioned in the introduction, finding the required dosage of VKAs  for a particular patient begins  just after surgery while he/she is in the hospital. The patient is discharged on this particular dose. Then the patient checks  PT-INR periodically and adjusts the dose initially every week for one month, then fortnightly for two months and then monthly thereafter.

\par There were many attempts using machine learning to predict the VKA dose requirement. Sharabiani et al \cite{ref_article1} developed a methodology to predict the initial dosage  of  warfarin  for  new  patients.  The  patients  were  classified  into 2 groups, those  who require more than 30mg of warfarin per week and those requiring less than 30mg per week. The classification was done using Relevance Vector Machines (RVM). For each class, customised regression models were developed to predict the dosage for each patient in the respective class. The dataset used in this approach was multi-ethnic and contained continuous variables  like  body surface area (BSA)  and PT-INR  along  with  categorical variables  such  as gender, race, presence of diabetes , and history of smoking with the most weightage  given to the BSA. The prediction accuracy was 11.6 in terms of root mean squared error (RMSE). This accuracy is insufficient to be practically used. One reason for high error rate is the high dependency of warfarin on the demographic data\cite{warfarin_demography}. Hence, using a multi-ethnic dataset to train the model could  result in lower accuracy.

\par Another  study by Schelleman et al \cite{ref_article2}, aimed  to develop a dosage prediction algorithm for Caucasians and African Americans by taking into account the clinical, environmental and genetic factors and comparing the result obtained with giving the empirical 5mg per day as maintenance dosage. The dataset considered variables like age, gender, body surface area and had information about the variants in CYP2C9 and VKORC1 genes which are responsible for the metabolism and action of the VKA drugs. Separate models were built for Caucasians and African Americans. In both the models, highest weightage was given to the VKORC1 gene variant variable. The model for Caucasians obtained better accuracy than the model for African Americans. The reason for this, they claimed, might be because they had not considered certain gene variants which might have been more important than the VKORC1 gene variant. Since the model focused primarily on Caucasians and African Americans, the same model may not be applicable to Indian population and we do not have genetic data from all our population.

 \par The members of the International  Warfarin Pharmacogenetics Consortium (IWPC) developed a pharmacogenetic prediction model \cite{ref_article4} to predict a stable therapeutic dose of warfarin and compared the result with that of a model which only considered the clinical factors and a model which gave fixed dosage to individuals. The dataset considered contained data of individuals from 9 countries and 4 continents whose target INR was between 2 and 3. Genotypic variables were taken into account along with the variables like age, race, height, etc. The model with the least predictive mean absolute error was chosen as the best model for all the three cases considered. They are 
 
 \begin{enumerate}
  \item Taking into account the pharmacogenetic factors,
  \item Taking into account only the clinical factors, and 
  \item Model which gave a fixed dose of 5mg of warfarin per day.
\end{enumerate}

    % \item Case 1: Taking into account the pharmacogenetic factors,
    % \item Case 2: Taking into account only the clinical factors, and 
    % \item Case 3: Model which gave a fixed dose of 5mg of warfarin per day.
    
 \par The performance of the algorithms was checked in three dose groups, those with less than 21 mg per week, those who require more than 49mg per week and those who require doses between 21 mg and 49 mg per week. The least squares linear regression modelling method was used to develop the required algorithm which gave the square root of the required dose. The result of this model claimed that the model taking into account the pharmacogenetic factors predicted the dosage with maximum accuracy. Then came the model which took into account the clinical factors.
 
 \par It was observed that in most of the cases, the physician predicts the dosage without considering the patient’s genotypic data, which is not usually available. The genetic factors influenced more in the initial dosage prediction, than the prediction of the maintenance dose. 
 
 \par Considering the above discussed factors, there  is a need for a better, more localised approach in developing the required algorithm. There are no predictive algorithms specific to Indian patients . We attempted to develop an algorithm using de-identified patient data obtained from Sree Chitra Tirunal Institute for Medical Sciences and Technology (SCTIMST).

\section{Data Collection, Analysis and Preprocessing}
\subsection{Dataset Collection}
This project is done in collaboration with National Institute of Technology Calicut (NITC) and SCTIMST. SCTIMST  provided the deidentified data of 109 Indian patients who are attending the INR Clinic of SCTIMST, to NITC. The data model parameters are described in Table \ref{table:dataParametersDescription}.

\begin{table}
\centering
\caption{Data Parameters description}
\begin{tabular}{ |p{1cm}|p{3cm}|p{6cm}|  }
\hline
S.No & Parameter & Description\\
\hline
1 & Age & Age of the patient \\
2 & Old INR Value   & INR value of patient before PT-INR test \\
3 & New INR Value  & INR value of patient after PT-INR test\\
4 & Old Dosage  & Old warfarin dosage of the patient (in mg)\\
5 & Gender & Gender of the patient \\
6 & Procedure type & Type of procedure the patient had undergone. It can be MVR, DVR, AVR or AF. \\
7 & New Dosage & New warfarin dosage prescribed by the doctor after the INR test (in mg)\\
\hline
\end{tabular}
\label{table:dataParametersDescription}
\footnotesize{\newline \newline  MVR - Mitral Valve Replacement, AVR - Aortic Valve Replacement, DVR - Double Valve Replacement, AF - Atrial Fibrillation, INR - International Normalised Ratio}
\end{table}

\subsection{Dataset Analysis}
A preliminary analysis was conducted on the dataset to figure out patterns and possible biases in the data. Frequency distribution of parameters are plotted to figure out shortcomings in dataset. The plots are given from Figure \ref{genderAnalysis} to Figure \ref{newINRAnalysis}.

\subsubsection{Gender Distribution}
The dataset has adequate representation from male and female population. Data points from other genders are absent, which may lead to inaccurate results for LGBTQ patients. See Figure \ref{genderAnalysis}.
\subsubsection{Procedure Type Distribution}
\begin{figure}
\centering
\begin{minipage}{.5\textwidth}
\centering
  \includegraphics[width = 65mm, height=5cm]{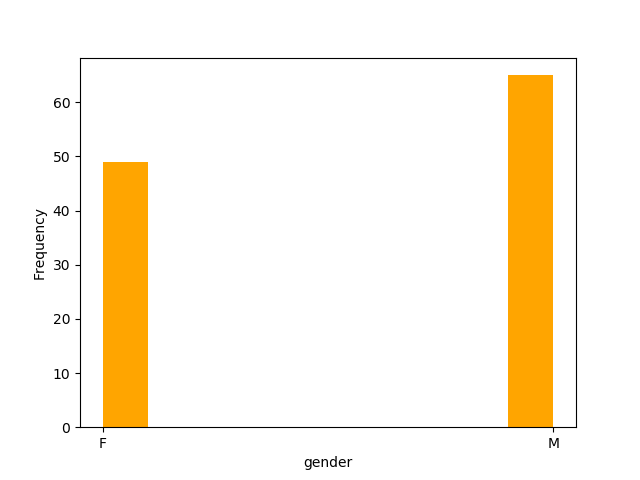}
  \captionof{figure}{Gender distribution}
  \label{genderAnalysis}
\end{minipage}%
\begin{minipage}{.5\textwidth}
\centering
  \includegraphics[width = 65mm, height=5cm]{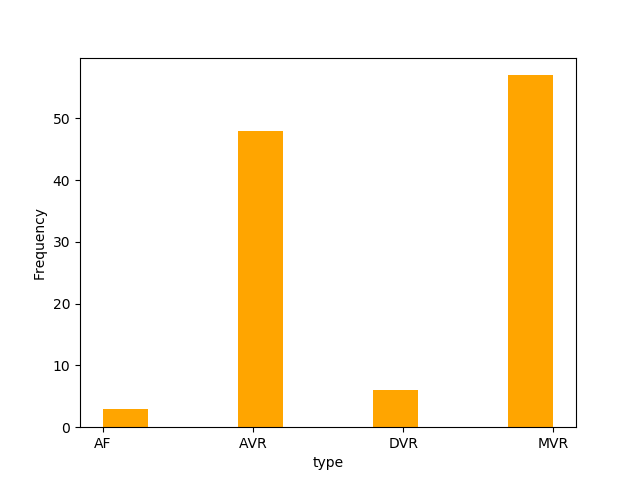}
  \captionof{figure}{Procedure type distribution}
  \label{procedureAnalysis}
\end{minipage}
\end{figure}

Patients undergoing AVR and MVR procedures are represented well. A shortage of data from AF and DVR categories can
be seen in Figure \ref{procedureAnalysis}.

\subsubsection{Age Distribution}
\begin{figure}
\centering
\begin{minipage}{.5\textwidth}
\centering
  \includegraphics[width = 65mm, height=5cm]{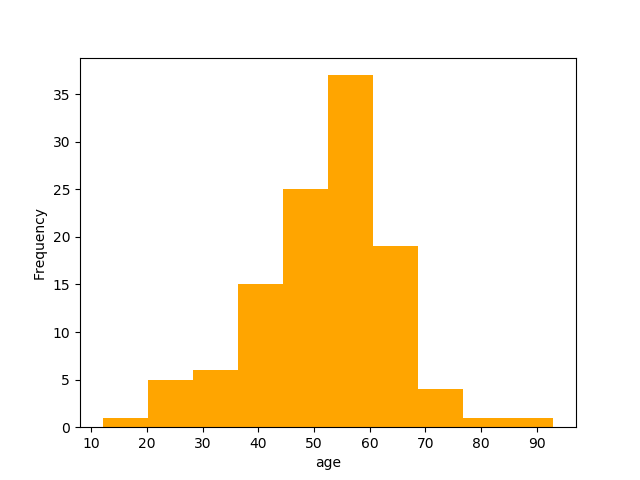}
  \captionof{figure}{Age distribution}
  \label{ageAnalysis}
\end{minipage}%
\begin{minipage}{.5\textwidth}
\centering
  \includegraphics[width = 65mm, height=5cm]{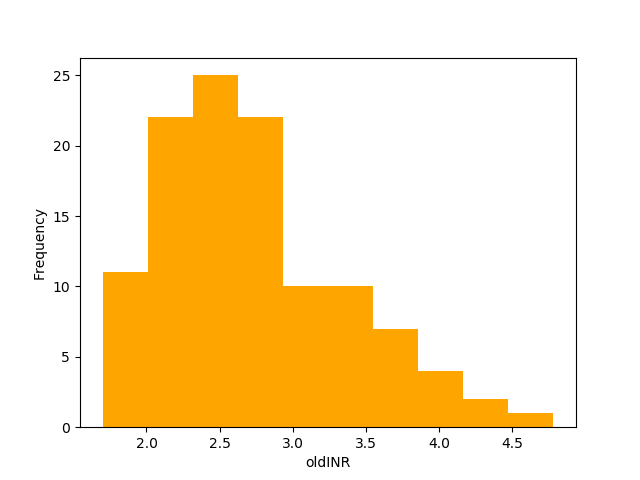}
  \captionof{figure}{Old INR distribution}
  \label{oldINRAnalysis}
\end{minipage}
\end{figure}

Patients from age 12 to 94 are present in the dataset. Patients in age group 40 to 70 are generally well represented. See Figure \ref{ageAnalysis}.

\clearpage

\subsubsection{Old INR value Distribution}
Adequate data points are present in the dataset with old INR values ranging from 1.7 to 5.5. See Figure \ref{oldINRAnalysis}.

\subsubsection{New INR Value Distribtion}
\begin{figure}
\centering
\begin{minipage}{.5\textwidth}
\centering
  \includegraphics[width = 65mm, height=5cm]{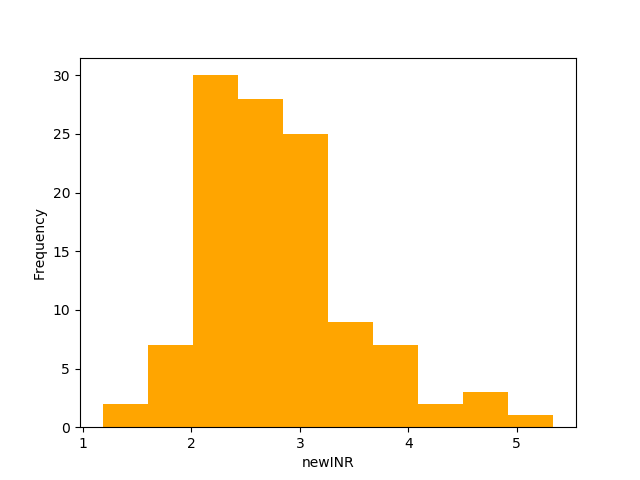}
  \captionof{figure}{New INR distribution}
  \label{newINRAnalysis}
\end{minipage}%
\end{figure}

Sufficient data points are present in the dataset with new INR values in target range 1.6 to 5.5. See Figure \ref{newINRAnalysis}.

\subsection{Data Preprocessing}
Data preprocessing is done to improve the accuracy of the model. The dataset has old dosage and new dosage given as daily dosage for some patients and as dosage sequence for other patients. The dosage sequence varies from fixed single daily dose to sequential dosing (two-day sequence to four-day sequence eg. 2 mg, 3 mg, 3 mg and this cycle repeats every third day). In order to attain uniformity in dataset, the sequence is averaged out to daily dosage by dividing the sum of sequence by the number of days.
\par Some patient data also contains dosage of acenocoumarol (acitrom), another commonly used VKA oral drug. Warfarin dosage is converted to acitrom dosage by dividing by a factor of two based on clinical data \cite{warfarin_switch}.
\par The categorical parameters present in the dataset (Gender and Procedure type) are encoded using one-hot encoding scheme. One hot encoding scheme encodes categorical parameters using binary representation to remove any extra weight assigned to higher integer value of categorical label in the schema \cite{one_hot}.  Gender is encoded with two binary variables and procedure type is encoded with four binary variables.

\section{Machine Learning Models}
A general procedure of randomly splitting dataset into 70\% training data and 30\% testing data was followed and a set of machine learning models was applied to detect accuracy.
\newline
The accuracy of regression models was compared using R square value.  \[R^2=(1-u/v)\] where, \[u=\sum_{n=1}^{dataset-size}(y(n)_{true}-y(n)_{predicted})^2\] 
\newline
 \[v=\sum_{n=1}^{dataset-size}(y(n)_{true}-y_{mean})^2\]
\newline
$y(n)$\textsubscript{true} = Real\ warfarin\ dosage\ of\ n\textsuperscript{th}\ patient
\newline
$y(n)$\textsubscript{predicted} = Predicted\ warfarin\ dosage\ of\ n\textsuperscript{th}\ patient
\newline

$y$\textsubscript{mean} = Mean\ real\ warfarin\ dosage 
\newline
The\ best\ possible\ value\ for\ R\ is\ 1.0.

\subsection{Linear Regression (LR) model}
In statistics, linear regression is a modelling tool used
for mapping the relationship between a scalar response and one or more explanatory variables.
The regression coefficients obtained in linear regression training is given in Table \ref{linear_regression}.

\begin{table}
    \centering
    \caption{Regression coefficients of LR Model}
    \begin{tabular}{ |p{3cm}|p{3cm}|  }
         \hline
        Feature &  Coefficient \\
        \hline
        Gender & .268093776 \\
        Procedure type & -.0392217685 \\
        Age & .000704527364 \\
        Old INR & .164906416 \\
        New INR & -.777297970 \\
        Old Dose & .917862769 \\
         \hline
    \end{tabular}
    \label{linear_regression}
\end{table}
\subsubsection{Observations}
A variance score of 0.951 and mean square error of  0.439 was observed which strongly suggests that data is linearly distributed. The highest coefficient was obtained for old dosage  and new INR suggesting that the target new dosage is more co-related to these parameters.

\subsection{Support Vector Regression (SVR) Model}
Support Vector Machines are models which are trained under the condition that an optimal hyperplane exists which separates the dataset into different classes \cite{svm}. It can be represented as the equation given below. 
\begin{equation}
f(x) = wx + b
\label{svm_eq}
\end{equation}
Where $f(x)$ is the optimal hyperplane with normal vector $w$ and intercept $b$.
Support vector regressions adds an extra condition that f(x) should satisfy 
\begin{equation}
    |f(x) - y(x) | \underline{<} e
\end{equation}
where y(x) is the labelled value function and e is the error tolerance.

\subsubsection{Kernels used}
Kernels can be used to map the dataset to different dimensions to obtain more seperability. The kernels tested include
\begin{enumerate}
    \item Linear kernel with e = 0.01
    \item Polynomial kernel with degree = 2 and e = 0.01
    \item Radial Basis Function kernel with gamma = 0.1 and e = 0.01
\end{enumerate}
%\newline
The optimal e value for training was found to be 0.01 through \textit{k fold cross validation}. The dataset was split into 10 groups (k = 10) in random and one group is chosen as the testing set and the rest as training set. Training is conducted with e values 0.001, 0.01, 0.1, 1.0, 10 and it is found that e = 0.01 gives better average variance.

\begin{table}
    \centering
    \caption{Coefficients of Linear SVMR Model}
    \begin{tabular}{ |p{3cm}|p{3cm}| }
         \hline
        Feature &  Coefficient \\
        \hline
        Sex & 0.95736124 \\
        Procedure & -0.00674263 \\
        Age & 0.13047947 \\
        Old INR & 1.36002697 \\
        New INR & -3.26086646 \\
        Old Dose & 0.955252 \\
         \hline
    \end{tabular}
    \label{svm_regression}
\end{table}

\subsubsection{Observations}
It is found that linear kernel model gives the lowest mean square error value of 0.41 and variance value of 0.955.
The coefficients of linear SVR are given in Table \ref{svm_regression}

\section{Weekly Dosage Prediction Algorithm}
The predicted daily dosage of warfarin has nanogram precision whereas the warfarin medicine is available only in doses of milligrams. Hence, an algorithm was designed to convert the daily decimal dosage (in mg) to a weekly sequence of integers. The algorithm takes predicted daily dosage and sequence length to produce best possible sequence with given sequence length and minimal error from predicted daily dosage. 
\begin{enumerate}
    \item Initialize lower-bound = floor(daily dosage)
    \item Initialize upper-bound = ceil(daily dosage)
    \item Initialize predicted-sequence = \{lower-bound\} with length sequence-length
    \item Initialize target-sum = predicted-dosage*sequence-length
    \item while(sum(predicted-sequence) - target-sum $<=$ 0) do
    \begin{itemize}
        \item previous-sequence = predicted-sequence
        \item replace last found occurrence of lower-bound with upper-bound in predicted-sequence
    \end{itemize}
    \item Return minimum of previous sequence, predicted-sequence with respect to the minimization function abs(sum(sequence) - target-sum)
\end{enumerate}

\section{Client Application}
The application was developed in the Android Platform. Since the app is going
to be used by the common man, care was taken to make it simple and user friendly. Initially, when the mobile application is loaded, the user has to authenticate with username and password. This will guide them to the main page. This main page will have fields namely patient's age, gender, old INR value, new INR value and old dosage, which the user has to enter. The initial entry is to be done by hospital staff and the patient needs to enter only the old INR value, new INR value and old dosage. The old dosage can be entered as a single value or a set of values. When the user clicks the Predict button, if the values entered by the user are all valid, the user is directed to the output page which shows the warfarin dosage of the patient for the next week. \\

The initial version of the application was tested in the heart failure clinic of SCTIMST and based on the feedback, many modifications were done in the application. The window to enter three consecutive day doses and the provision the enter the drug acenocoumarol (acitrom) were added. The conversion from acitrom to warfarin is by mutiplying with a factor of 2 as the drug is more potent. The current version of the app, displays the drug doses for a week starting from the date of entry not in descending order as before (eg 3mg, 3mg, 2mg, 2mg, 2mg, 2mg, 2mg) but varying doses interspersed(eg 2mg, 3mg, 2 mg, 3mg, 2mg, 2mg, 2mg)

\begin{figure}
\centering
\begin{minipage}{.33\textwidth}
\centering
  \includegraphics[scale=0.07]{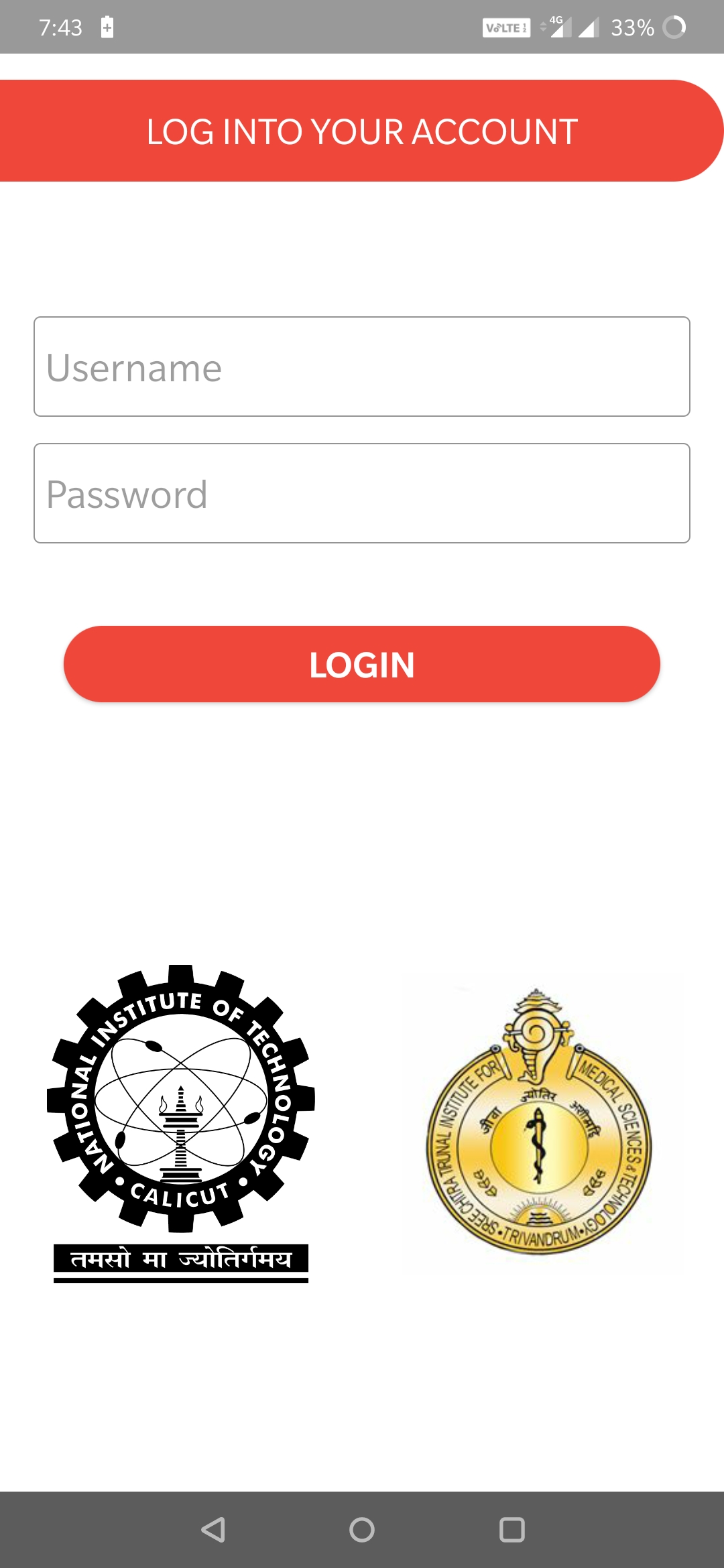}
  \captionof{figure}{Login Screen}
  \label{loginscreen}
\end{minipage}%
\begin{minipage}{.33\textwidth}
\centering
  \includegraphics[scale=0.07]{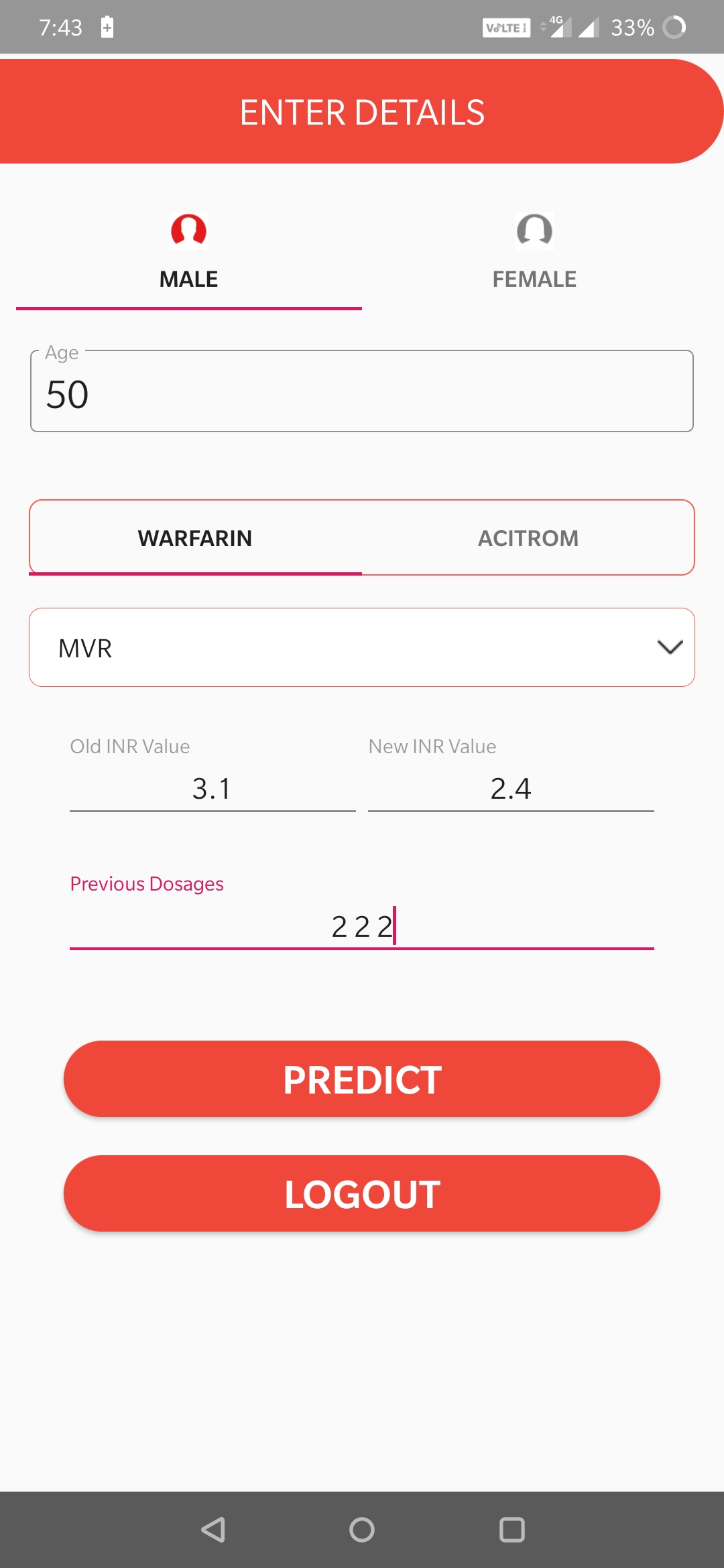}
  \captionof{figure}{Input Screen}
  \label{inputscreen}
\end{minipage}%
\begin{minipage}{.33\textwidth}
\centering
  \includegraphics[scale=0.07]{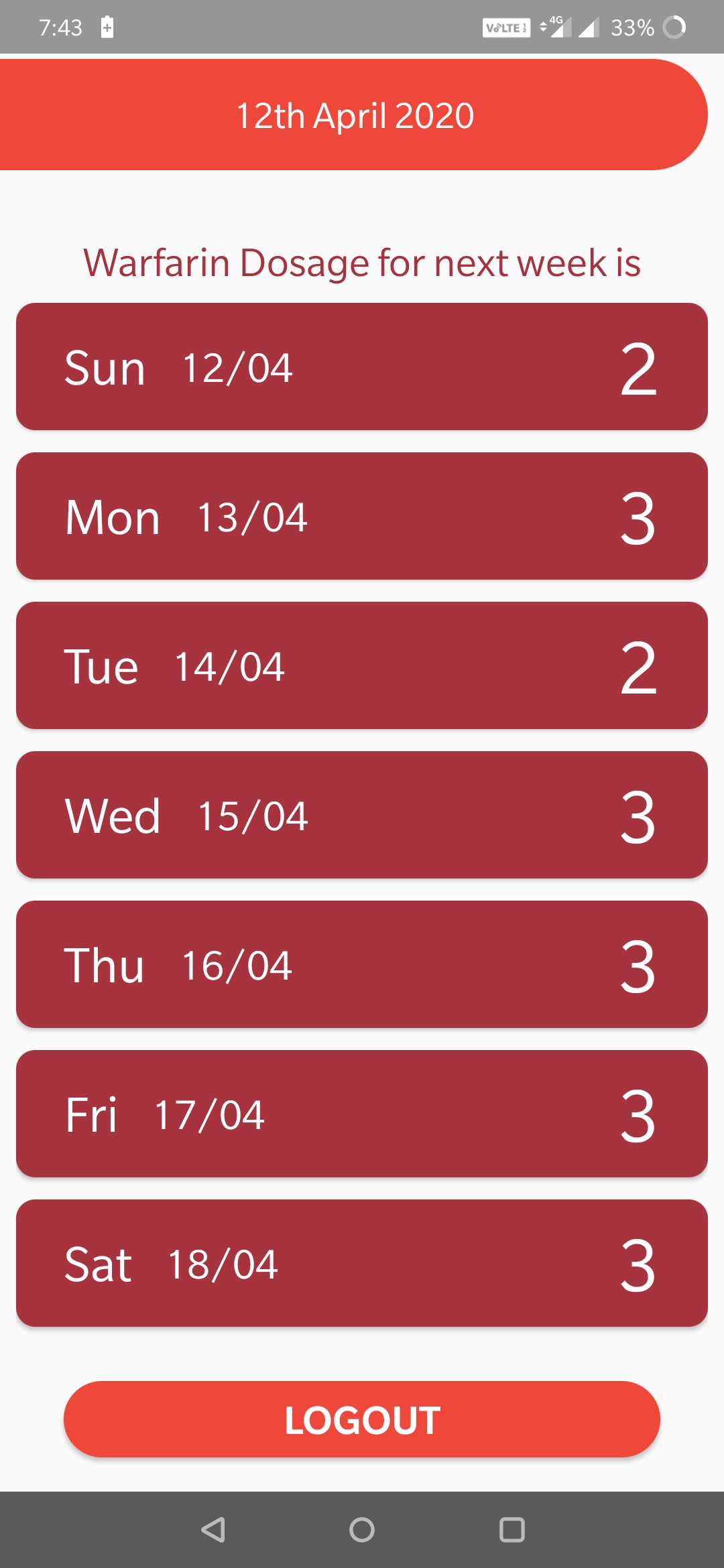}
  \captionof{figure}{Output Screen}
  \label{outputscreen}
\end{minipage}
\end{figure}

\section{Results}
Various machine learning models are compared and linear SVM Regression model with coefficients given in Table \ref{svm_regression} gave the best variance score and accuracy. Regression models with linear base generally outperforms other models in prediction. It is found out that new INR value and old dosage are the principal components in prediction of new dosage. The daily dosage predicted by the server is stable and usable in case of stable INR value range of two to four. For an INR value below 2 and above 4,  it was decided to refer for the help of a physician as he/she may need urgent medical help. The weekly prediction model also provides reasonable accuracy with one-off errors in boundary cases. Weekly model is going to be tested in the INR clinic of SCTIMST for further improvement in accuracy. A variance score of 0.955 was obtained for daily prediction with mean square error of 0.41.

\section{Pre testing the clinical accuracy of the application}
The application was pretested using a different set of 50 physician assigned values for INR prediction in the INR Clinic of SCTIMST. It was found that the application predicted the values accurately in lower INR ranges, but towards the higher range(3.5-4 range of INR), there was variation compared to physician assigned values in the tune of upto 5 mg in weekly doses. This
indicates the need for further refinement of the algorithm, probably by re-training the algorithm with more physician derived datasets. We are planning to use 500 more data to try and improve the
accuracy of the algorithm.

\section{Future Research - Clinical testing of the Algorithm in patients}
Once we improve the accuracy as described above and find the values in the acceptable range compared to physician derived values, we will approach the ethics committee of SCTIMST to test the efficacy and utility of this application in patients attending the INR clinic of SCTIMST. Once the efficacy and utility is proven it will be released to the public.

\section{Conclusion}
\par A warfarin dosage prediction algorithm was developed using data from Indian patients. The linear regression model and the support vector regression model were tested and the support vector regression model was found to show better results with a lowest mean square error value of 0.41 and a variance value of 0.955. A mobile application was developed using the algorithm and is going to be tested in the INR clinic of SCTIMST in a larger group of patients for ease of use and accuracy. The application, after testing, can be used for prediction of daily and weekly dosage of warfarin and acenocoumarol for patients without consulting a physician. Patients who are on the above oral anticoagulants from remote areas can use this application installed in their mobile phones.
%\bibliographystyle{splncs04}
\bibliography{sample}

\end{document}